\documentclass[12pt, amsmath, amssymb, showpacs, nofootinbib, sort&compress, showkeys, round, citeautoscript]{revtex4}

\setcitestyle{super}

\usepackage{graphics}
\usepackage{epsfig}
\usepackage{subfigure}  
\usepackage{bm}
\usepackage{booktabs}       
\usepackage{threeparttable} 

\makeatletter
\def\@biblabel#1{(#1)}
\makeatother

\begin{document}
\title{Modeling of Lead Halide Perovskites for Photovoltaic Applications}

\author{
Radi A. Jishi*$^1$, Oliver B. Ta$^2$, Adel A. Sharif$^3$}
\affiliation{$^1$ Department of Physics, California State University, Los Angeles, CA\\
$^2$ Department of Mechanical Engineering, California State University, Los Angeles, CA}

\date{\today}

\begin{abstract}
We report first-principles calculations, using the full potential linear augmented plane wave method, on six lead halide semiconductors, namely, CH$_3$NH$_3$PbI$_3$, CH$_3$NH$_3$PbBr$_3$, CsPbX$_3$ (X$=$Cl, Br, I), and RbPbI$_3$.  Exchange is modeled using the modified Becke-Johnson potential.  With an appropriate choice of the parameter that defines this potential, an excellent agreement is obtained between calculated and experimental band gaps of the six compounds.  We comment on the possibility that the cubic phase of CsPbI$_3$, under hydrostatic pressure, could be a topological insulator.
 
\end{abstract}


\maketitle

\section{I. INTRODUCTION}\label{sec:1}

Recently, materials with halide perovskite structure, with the general formula ABX$_3$, have attracted great interest, primarily because of their potential applications as light harvesters in solar cells \cite{ref:Kojima} and as topological insulators.\cite{ref:Jin,ref:Yang}  Many studies have ensued with the aim of both improving the performance of these materials in photovoltaic cells and of understanding which physical parameters may determine the efficiencies.\cite{ref:Etgar, ref:Ball, ref:Heo, ref:Kim, ref:Bi, ref:Cai, ref:Eperon, ref:Laban, ref:Stranks, ref:Mosconi, ref:Wang, ref:GW, ref:Lee, ref:Noh, ref:Burschka, ref:Liu}  For example, Lee et al.\cite{ref:Lee} report a solution-processable solar cell which uses a perovskite of mixed halide form, namely methylammonium lead iodide chloride, CH$_3$NH$_3$PbI$_2$Cl (abbreviated as MAPbI$_2$Cl), with a solar-to-electrical power conversion efficiency of 10.9\%.  Using chemically-tuned MAPb(I$_{(1-x)}$Br$_x$)$_3$ perovskites as light harvesters, a mesoporous titanium dioxide (TiO$_2$) film, and a hole-conducting polymer, Noh et al. \cite{ref:Noh} demonstrate solar cells with a 12.3\% power conversion efficiency.  Burschka et al. \cite{ref:Burschka} describe a sequential deposition method whereby MAPbI$_3$ nanoparticles are formed within porous TiO$_2$, resulting in a power conversion efficiency of 15\%.  Liu et al. \cite{ref:Liu} have subsequently shown that such nanostructuring  is not necessary for high efficiencies; a planar heterojunction solar cell, with a deposited thin film of MAPbI$_2$Cl acting as a light absorber, can achieve an efficiency exceeding 15\%.  

MAPbI$_3$ and similar compounds are derived from a class of trihalide perovskite structures with the formula ABX$_3$ (A=Cs, Rb; B=Pb; X=Cl, Br, I) by replacing the alkali-metal atom with methylammonium (MA).  Such a replacement causes a large downshift in the semiconducting energy gap, making the compounds useful for photovoltaic applications.  It is anticipated that different band gaps may be obtained by replacing methylammonium with other entities such as NH$_4$ or  CH$_2$CH, by applying pressure, or by using thin films consisting of only a few layers.

To maximize the usefulness of such materials in photovoltaic applications, it is important to begin by developing computational techniques that accurately describe their electronic structure.  Density functional theory in the Kohn-Sham formulation \cite{ref:Kohn} is the most widely-used method.  Here, the exchange-correlation potential is approximated by a functional of the electronic density.  The most common approximations are the local density approximation (LDA),\cite{ref:Kohn} the generalized gradient approximation (GGA),\cite{ref:Perdew} and the hybrid approximation.\cite{ref:Becke93}  While LDA and GGA provide a successful description of ground-state properties in crystals, this success does not extend to a description of excited states.  In many semiconductors, LDA and GGA strongly underestimate the value of the energy gap.  Improved values for the band gaps are usually obtained by using the GW method.\cite{ref:Bechstedt}  However, the high computational cost of this method limits its applicability to crystals with a small number of atoms in the unit cell.

An exchange potential was recently proposed by Becke and Johnson (BJ), designed to yield the exact exchange potential in atoms.\cite{ref:Becke06}  Unfortunately, the use of this potential led to a slight improvement in the energy gap values for many semiconductors.\cite{ref:Tran07}  A simple modification of the BJ potential was proposed by Tran and Blaha.\cite{ref:Tran09}  In this method, known as TB-mBJ, the exchange potential is given by

\begin{equation}
  V_x^{TB\_mBJ}(\mathbf{r})=cV_x^{BR}(\mathbf{r})+(3c-2)\frac{1}{\pi}\sqrt{\frac{5}{12}}[2t(\mathbf{r})/\rho(\mathbf{r})]^{1/2}
\label{eq:tbmbjpot}
\end{equation}
where 

\begin{equation}
  \rho(\mathbf{r})=\sum\limits_{i=1}^N|\psi_i(\mathbf{r})|^2
\end{equation}
is the electron density ($N$ is the number of occupied orbitals and $\psi_i$ is the Kohn-Sham (KS) $i^{th}$ orbital wave function),

\begin{equation}
  t(\mathbf{r})=\frac{1}{2}\sum\limits_{i=1}^{N}[\bm{\nabla}\psi_i^*(\mathbf{r})\cdot\bm{\nabla}\psi_i(\mathbf{r})]
\end{equation}
is the KS kinetic energy density, and 

\begin{equation}
  V_x^{BR}(\mathbf{r})=-\frac{1}{b(\mathbf{r})}\Big[1-e^{-x(\mathbf{r})}-\frac{1}{2}x(\mathbf{r})e^{-x(\mathbf{r})}\Big]
\end{equation}
is the Becke-Roussel exchange potential.\cite{ref:Becke89}  The function $x(\mathbf{r})$ in the above equation is determined by a nonlinear equation involving $\rho$, $\bm{\nabla}\rho$, $\bm{\nabla}^2\rho$, and $t$.  Once $x(\mathbf{r})$ is found, $b(\mathbf{r})$ is determined by 

\begin{equation}
  b=x[e^{-x}/(8\pi\rho)]^{1/3}
\end{equation}
In the TB-mBJ potential given in Eq~(\ref{eq:tbmbjpot}), 

\begin{equation}
  c=A+B\sqrt{g}
\end{equation}
where

\begin{equation}
g=\frac{1}{\Omega}\int\frac{1}{2}\bigg(\frac{|\bm{\nabla}\rho_\uparrow(\mathbf{r})|}{\rho_\uparrow(\mathbf{r})}+\frac{|\bm{\nabla}\rho_\downarrow(\mathbf{r})|}{\rho_\downarrow(\mathbf{r})}\bigg)d^3r
\end{equation}
is the average of $|\bm{\nabla}\rho/\rho|$ over the unit cell of volume $\Omega$.  The parameters $A=-0.012$ and $B=1.023$ bohr$^{1/2}$ were chosen because they produce the best fit to the experimental band gaps of many semiconductors.  Studies have shown that the TB-mBJ potential is generally as accurate in predicting the energy gaps of many semiconductors as the much more expensive GW method.\cite{ref:Koller11}

Despite its many successes, however, the performance of the TB-mBJ method is not very satisfactory in certain cases, especially for transition metal oxides.  To improve the band gap prediction, Koller, Tran, and Blaha \cite{ref:Koller12} consider a more general form for c,

\begin{equation}
c=A+Bg^e
\end{equation}
They vary the values of parameters $A$, $B$, and $e$ in order to improve the quality of the fit between the calculated and the experimental energy gaps of many semiconductors.  There is an overall improvement in predicting the energy gaps of semiconductors with moderate gaps when $A=0.267$, $B=0.656$, and $e=1$.  The modified BJ method employing these values for $A$, $B$, and $e$ will be referred to as the KTB-mBJ method.  It should be pointed out that, in terms of computational time and resources, the requirements for the TB-mBJ and KTB-mBJ methods are essentially the same as those for standard LDA or GGA methods.  Therefore, these methods may be easily used to calculate the electronic structure of crystals with large unit cells, where the cost of the GW method is prohibitive.

In this work, we present first-principles calculations on the electronic structure of six compounds, namely MAPbI$_3$, MAPbBr$_3$, CsPbX$_3$ (X=Cl, Br, I), and RbPbI$_3$.  All of these compounds have a perovskite structure, characterized by a Pb atom that is octahedrally coordinated to six halogen atoms.  We show that GGA, when spin-orbit coupling (SOC) is included, severely underestimates the band gaps in these semiconducting materials.  Though TB-mBJ and KTB-mBJ methods lead to significant improvement in the values of the gaps, both methods still underestimate the energy gaps by a wide margin.  We then show that keeping parameters $B$ and $e$ essentially the same as in TB-mBJ, while adopting a new value for $A$ leads to results that are in excellent agreement with experimental values.

\section{II. METHODS}\label{sec:2}

Total energy and band structure calculations are carried out using the all-electron, full potential, linear augmented plane wave (FP-LAPW) method as implemented in the WIEN2k code.\cite{ref:Blaha}  Here, each atom is surrounded by a muffin-tin sphere, and the total space is divided into two regions.  One region consists of the interior of these nonoverlapping spheres, while the rest of the space constitutes the interstitial region.  The radii of the muffin-tin spheres are 2.5 $a_0$ for Cs, Rb, Pb, I, and Br, 2.37 $a_0$ for Cl, 1.27 $a_0$ for N, 1.33 $a_0$ for C, and 0.68 $a_0$ for H, where $a_0$ is the Bohr radius.  In GGA calculations, the exchange correlation potential is that proposed in reference.\cite{ref:Perdew}  

The valence electrons' wave functions inside the muffin-tin spheres are expanded in terms of spherical harmonics up to $l_{max}=10$.  In the interstitial regions, they are expanded in terms of plane waves, with a wave vector cutoff of $K_{max}$.  Because of the small muffin-tin radius of hydrogen atoms, we set $R_{H}K_{max}=3$ in CH$_3$NH$_3$PbI$_3$ and CH$_3$NH$_3$PbBr$_3$, where $R_H=0.68\ a_0$ is the muffin-tin radius of the H atom.  In the remaining four compounds, we set $R_{mt}K_{max}=9$, where $R_{mt}$ is the smallest muffin-tin radius.  The charge density is Fourier expanded up to a maximum wave vector of $G_{max}$, where $G_{max}=20$ $a_0^{-1}$ for MAPbI$_3$ and MAPbBr$_3$, and $G=13$ $a_0^{-1}$ for the remaining compounds.  The convergence of the self-consistent calculations is achieved with a total energy tolerance of 0.1 mRy and a charge convergence of 0.001 e. 

\section{III. RESULTS AND DISCUSSION}\label{sec:3}

\begin{figure}[ht]
\includegraphics[height=6cm]{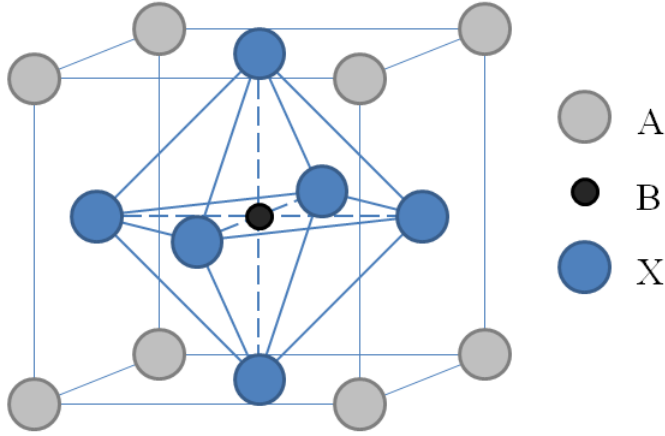}
\caption{(Color online) Cubic perovskite structure with alkali atoms occupying the A sites, Pb atoms occupying the B sites, and halogen atoms occupying the X sites.\label{fig:1}}
\end{figure}

At high temperatures, lead halide perovskites have a simple cubic unit cell, where Pb sits at the center of the cube and is octahedrally coordinated to six halogen atoms, while alkali atoms sit at the cube corners, as shown in Fig.~\ref{fig:1}.  As the temperature is lowered, distortions lead to tetragonal and/or orthorhombic structures.  In our calculations, we use the room temperature crystal structures of the various compounds.  These are listed in Table~\ref{tab:1}. For CsPbI$_3$, which is orthorhombic at room temperature, we also study the high temperature cubic phase, which was predicted to be a topological insulator when subjected to hydrostatic pressure.\cite{ref:Jin}

\begin{table}[ht]
  \caption{Crystal structure and lattice constants for the compounds studied in this work.}
    \begin{threeparttable}
      \begin{tabular}{ccc}
	\hline
	\hline
	Compound & Structure & Lattice constants (\AA)\\
	\hline
	CH$_3$NH$_3$PbI$_3$         & Tetragonal	& a=8.856, c=12.655$^a$			\\
	CH$_3$NH$_3$PbBr$_3$        & Cubic		& a=5.933$^b$				\\
	CsPbCl$_3$                  & Cubic		& a=5.605$^c$				\\
	CsPbBr$_3$                  & Orthorhombic	& a=8.244, b=11.735, c=8.198$^d$	\\
	RbPbI$_3$                   & Orthorhombic	& a=10.276, b=4.779, c=17.393$^e$	\\
	CsPbI$_3$                   & Orthorhombic	& a=10.458, b=4.801, c=17.776$^e$	\\
	CsPbI$_3$		    & Cubic		& a=6.2894$^e$				\\
	\hline
	\hline
      \end{tabular}
      \begin{tablenotes}
       \item[a] Poglitsch and Weber \cite{ref:Poglitsch}	\\
       \item[b] Mashiyama {\it et al.} \cite{ref:Mashiyama}	\\
       \item[c] Moreira and Dias \cite{ref:Moreira}		\\
       \item[d] Stoumpos {\it et al.} \cite{ref:Stoumpos}	\\
       \item[e] Trots and Myagkota \cite{ref:Trots}		\\
      \end{tablenotes}
      \label{tab:1}
    \end{threeparttable}
\end{table}

We carried out band structure calculations on the six compounds shown in Table \ref{tab:1}.  Using GGA, we found that, upon including the effect of SOC, the band gaps of all compounds are severely underestimated.  The values of the gaps are improved by using the TB-mBJ method, and are improved further by using the KTB-mBJ method.  However, the improvement does not go far enough, and the gaps are still far below the experimental values.  We thus considered a new set of values for the parameters that appear in Eq. (8), namely,

\begin{equation}
A=0.4,\:B=1.0,\:e=0.5 
\end{equation}

The parameters $B$ and $e$ are essentially the same as in the TB-mBJ method, but the parameter $A$ is different.  With this new set of values for $A$, $B$, and $e$, the calculated band gaps of all six compounds are in excellent agreement with the experimental values.  Our results are summarized in Table~\ref{tab:2} where we present the band gaps calculated by using different methods.  The gaps obtained by using the above values for $A$, $B$, and $e$ are given in the column labeled 'Present method.'

\begin{table*}[t]  
  \caption{Calculated and experimental band gaps, in eV, for the compounds that are studied in this work.  The band gaps obtained by using parameters in Eq. (9) are reported under ``Present method.''}
    \begin{threeparttable}
      \begin{tabular}{ccccccc}
	\hline
	\hline
	Compound 		    & GGA 	& GGA+so	& TB-mBJ 	& KTB-mBJ 	& Present method	& Experimental		\\
	\hline
	CH$_3$NH$_3$PbI$_3$         & 1.492	& 0.377		& 0.844 	& 0.921 	& 1.544 		& 1.5-1.6$^{[a,b]}$	\\
	CH$_3$NH$_3$PbBr$_3$        & 1.668	& 0.453 	& 1.183 	& 1.406 	& 2.233 		& 2.28$^{[a]}$  	\\
	CsPbCl$_3$                  & 2.498	& 0.707 	& 1.585 	& 1.889 	& 2.829 		& 2.86$^{[c]}$  	\\
	CsPbBr$_3$                  & 1.794	& 0.669 	& 1.316 	& 1.461 	& 2.228 		& 2.24$^{[c,d]}$  	\\
	RbPbI$_3$                   & 2.468	& 1.828 	& 2.387 	& 2.446 	& 3.302 		& 3.17$^{[e]}$  	\\
	CsPbI$_3$                   & 2.504	& 1.876 	& 2.426 	& 2.476 	& 3.330 		& 3.14$^{[e]}$  	\\
	CsPbI$_3$ (cubic)	    & 1.324	& 0.072 	& 0.485 	& 0.529 	& 1.072			& -			\\
	\hline
	\hline
      \end{tabular}
      \label{tab:2}
      \begin{tablenotes}
       \item[a] Noh {\it et al.} \cite{ref:Noh}			\\
       \item[b] Baikie {\it et al.} \cite{ref:Baikie}		\\
       \item[c] Liu {\it et al.} \cite{ref:Liu13}		\\
       \item[d] Stoumpos {\it et al.} \cite{ref:Stoumpos}	\\
       \item[e] Yunakova {\it et al.} \cite{ref:Yunakova}	\\
      \end{tablenotes}
  \end{threeparttable}
\end{table*}

\begin{figure}[htbp]
  \includegraphics[height=8.5cm]{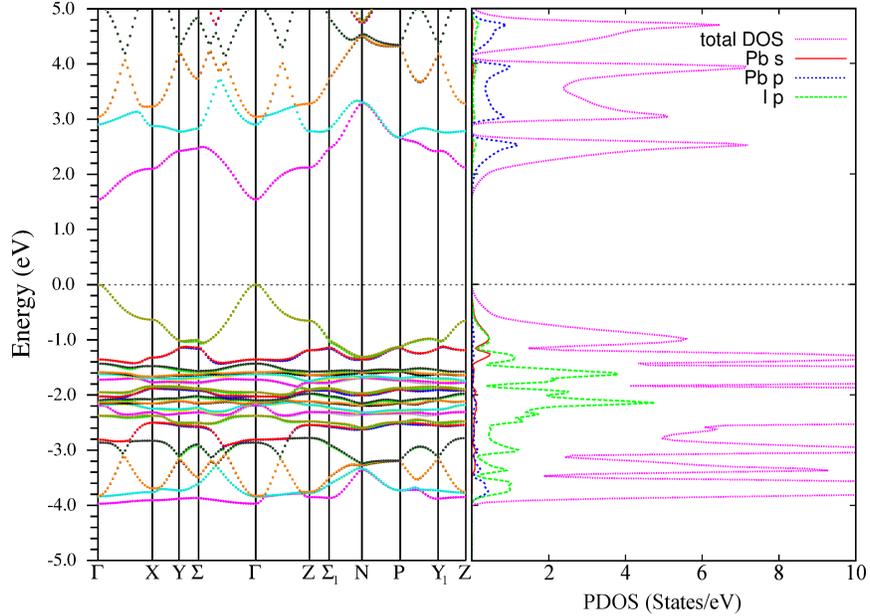}
  \caption{(Color online) Band structure and density of states of CH$_3$NH$_3$PbI$_3$.\label{fig:2}}
\end{figure}

The calculated energy bands of MAPbI$_3$ along high symmetry directions in the Brillouin zone (BZ), in addition to the electronic density of states, are presented in Fig.~\ref{fig:2}.  The valence band maximum (VBM) and conduction band minimum (CBM) occur at the $\Gamma$-point, the BZ center.  In cubic perovskites, the gap occurs at point R(1/2, 1/2, 1/2). However, at room temperature, MAPbI$_3$ has a body-centered tetragonal crystal structure with two formula units per primitive cell.  Its conventional unit cell, containing four formula units, is a slightly distorted $\sqrt{2}\times\sqrt{2}\times2$ supercell of the high temperature cubic phase unit cell.  The distortion consists mainly of a rotation of the octahedron by $10.45\,^{\circ}$ about the c-axis.  Point R of the cubic lattice BZ is zone-folded into the $\Gamma$-point of the body-centered tetragonal lattice BZ.

\begin{figure*}[t]
  \includegraphics[height=8.5cm]{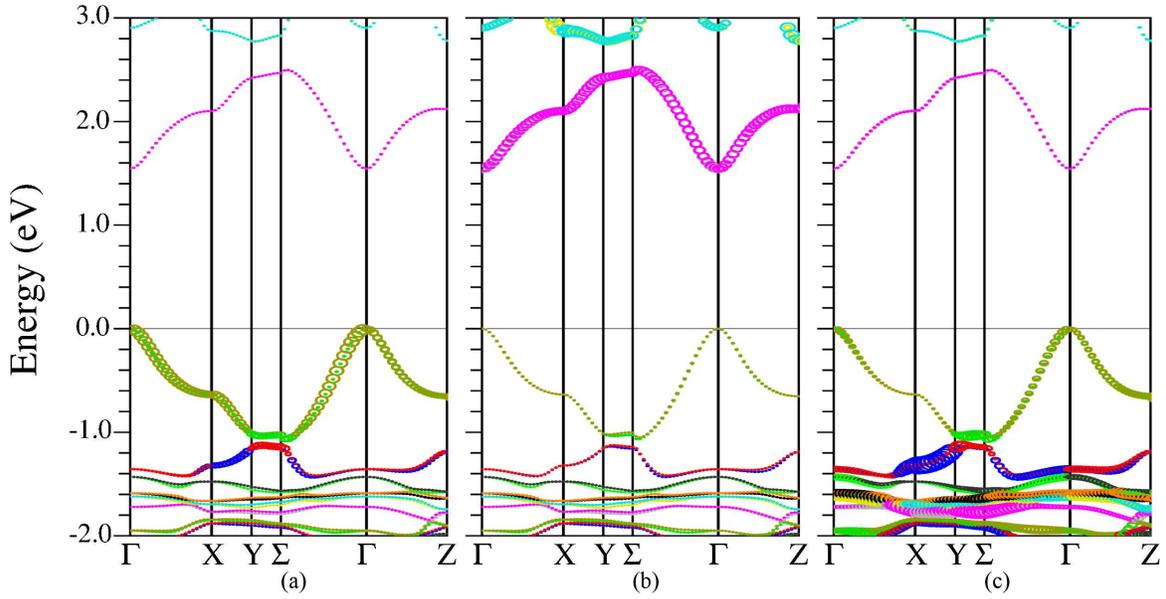}
  \caption{(Color online) Orbital character of the valence and conduction bands of CH$_3$NH$_3$PbI$_3$.  The contribution of the selected orbital is proportional to the size of the circle, with a single point denoting zero contribution. (a) Pb 6s orbital, (b) Pb 6p orbital, and (c) I 5p orbital.
  \label{fig:3}}
\end{figure*}

The density of states of MAPbI$_3$ is shown in Fig.~\ref{fig:2}, where we see that the low-lying conduction bands are derived from Pb p states.  On the other hand, the bands in the range -4 eV to -2 eV are dominated by iodine-derived states.  The valence band just below the Fermi energy is derived from lead s and iodine p states.  These observations become clear upon considering the atomic orbital character of the bands, which is presented in Fig.~\ref{fig:3}.  The size of the circles is indicative of the contribution of the chosen atomic orbital to the eigenstates at each \textbf{k}-point.  The CBM is derived mostly from Pb 6p states.  The VBM, on the other hand, is a mixture of Pb 6s and I 5p states.  The antibonding state formed from these s and p states is pushed up in energy close to the Fermi level.  The large contribution of Pb 6s $(l=0)$ states to the VBM and Pb 6p $(l=1)$ states to the CBM suggests that there are strong optical transitions between the VBM and CBM $(\Delta l=1)$, hence the usefulness of this material in solar applications.  

\begin{figure}[htbp]
  \includegraphics[height=8.5cm]{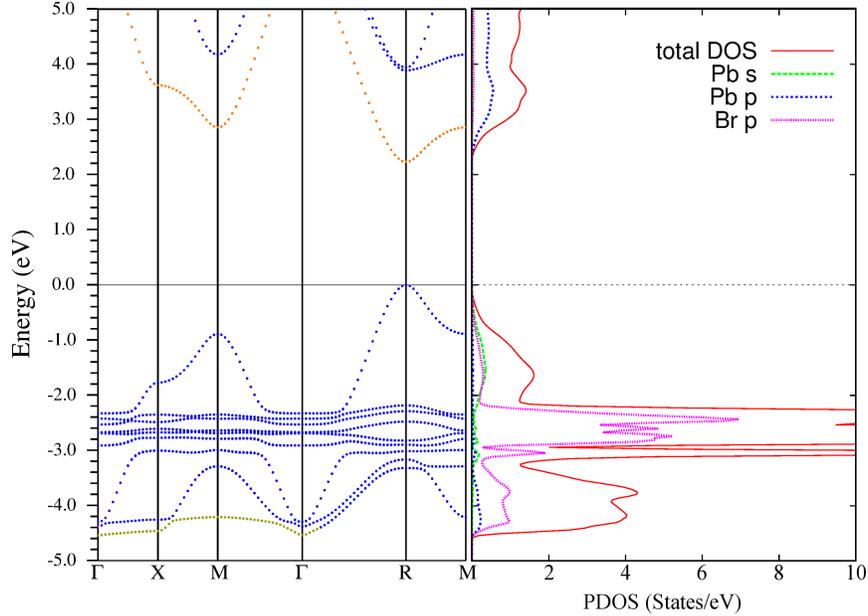}
  \caption{(Color online) Band structure and partial density of states of CH$_3$NH$_3$PbBr$_3$.\label{fig:4}}
\end{figure}

CH$_3$NH$_3$PbBr$_3$ (MAPbBr$_3$) has a cubic unit cell.  Its band structure is shown in Fig.~\ref{fig:4}, and, as expected, the band gap occurs at point R(1/2, 1/2, 1/2) in the Brillouin zone.  As with the case of MAPbI$_3$, its VBM is a mixture of Pb 6s and Br 4p states, whereas its CBM is derived from Pb 6p states.  In the absence of SOC, its CBM is six-fold degenerate (including spin degeneracy).  Due to SOC, its CBM is split into a doublet $(j=1/2)$ and a quartet $(j=3/2)$.  The doublet is lowered in energy by an amount $\lambda$, whereas the quartet is raised in energy by ${\lambda}$/${2}$, where $\lambda\approx1.1$ eV.  Similar perovskite structures, namely CsSnX$_3$ (X=Cl, Br, I), where Sn replaces Pb, show a much smaller spin splitting of $\sim0.4$ eV.\cite{ref:Lambrecht}  Since the VBM is composed of Pb s and Br p orbitals, it is shifted slightly upward due to SOC on Br atoms.  The large energy split of the CBM is, of course, due to the strong SOC on Pb atoms. 

Finally, we consider the CsPbI$_3$ crystal.  At high temperature ($>$634K), the crystal is cubic, but at room temperature, it is strongly distorted to an orthorhombic structure.  Based on LDA and sx-LDA calculations, it has been suggested that, under hydrostatic pressure, the cubic phase might become a topological insulator.\cite{ref:Jin}  Calculations made using sx-LDA suggest a gap of 0.566 eV for CsPbI$_3$ and 0.218 eV for CsSnI$_3$.  With decreasing lattice constants, the band width increases and the band gap decreases; at some critical pressure, band inversion occurs.  For CsPbI$_3$ and CsSnI$_3$, those critical pressures are predicted to be 3.33 GPa and 0.96 GPa, respectively.  However, GW calculations on CsSnI$_3$ give a much larger band gap of 1.008 eV.\cite{ref:Lambrecht}  Our calculation on the cubic phase of CsPbI$_3$ shows that the band gap is 1.07 eV, larger  by 0.5 eV than predicted by sx-LDA.  Assuming linear dependence of the energy gap on lattice constant,\cite{ref:Jin} a critical pressure of ~6.6 GPa has to be applied to cause band inversion. 

\begin{figure}[htbp]
  \includegraphics[height=8.5cm]{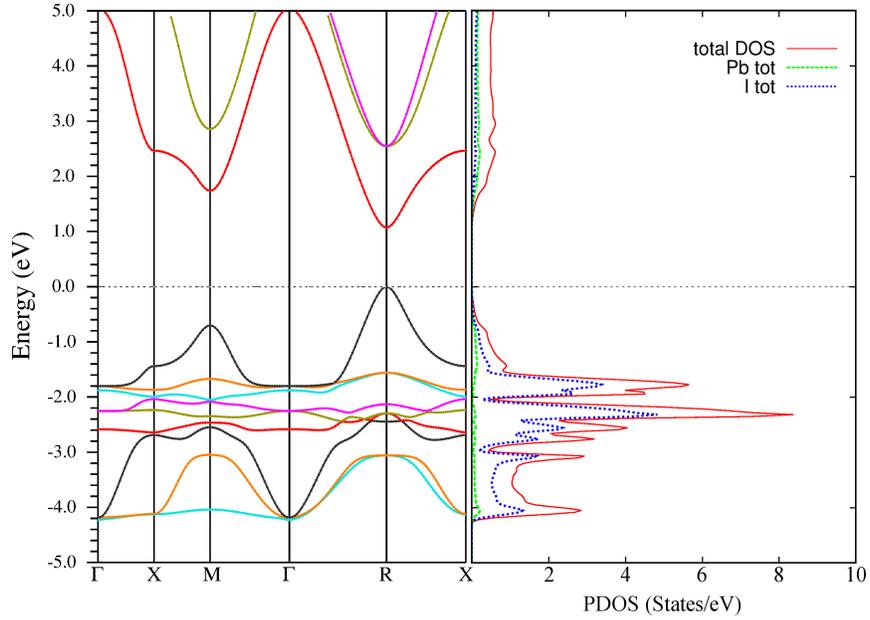}
  \caption{(Color online) Band structure and partial density of states of cubic CsPbI$_3$.\label{fig:5}}
\end{figure}

The band structure of cubic CsPbI$_3$ is shown in Fig.~\ref{fig:5}.  The band gap occurs at point R.  As in the cases discussed previously, its CBM is derived from Pb 6p states, whereas its VBM is a mixture of Pb 6s and I 5p states.  Without SOC, the calculated band gap is 2.27 eV, and the CBM is six-fold degenerate (including spin degeneracy).  SOC on Pb splits its CBM into a doublet $(j=1/2)$ and a quartet $(j=3/2)$.  The doublet is lowered in energy by 1.1 eV, while the quartet is raised by 0.55 eV.  On the other hand, SOC on the I atoms raises the VBM by 0.1 eV.

\begin{figure}[htbp]
  \includegraphics[height=8.1cm]{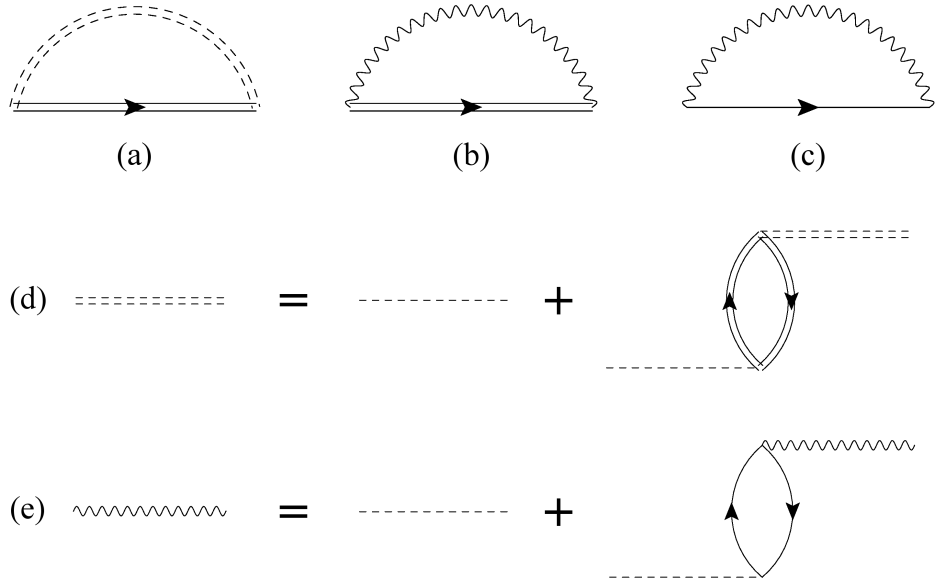}
  \caption{The electron's proper self energy in the (a) GW approximation, (b) GW$_0$ approximation, and (c) G$_0$W$_0$ approximation.  W and W$_0$ are given in (d) and (e), respectively.  The single solid line is the noninteracting electron propagator, while the double solid line is the interacting electron propagator.  The single dashed line is the bare Coulomb interaction.  The double-dashed line (W) is the screened Coulomb interaction in the GW approximation, while the wavy line (W$_0$) is the screened Coulomb interaction in the random phase approximation.\label{fig:6}}
\end{figure}

As a further check on our results for cubic CsPbI$_3$, we repeated the calculation of the band gap using the GW approximation within the linear augmented plane wave formalism.\cite{ref:gap}  In this method, the electron's proper self energy $\Sigma$* is approximated as a product of the electron Green's function (G) and an effective interaction term (W).  We carried out the calculation in the absence of spin-orbit coupling and using the G$_0$W$_0$ and GW$_0$ approximations.  The electron's proper self energy in these approximations is shown graphically in Fig.~\ref{fig:6}.  We found that within the G$_0$W$_0$ approximation, the band gap is 2.04 eV, and it increases to 2.19 eV upon employing the GW$_0$ approximation.  This result is in excellent agreement with, the value of 2.27 eV, which we obtained for the band gap, in the absence of SOC, by using the modified Becke-Johnson form of the exchange potential.

In conclusion, we have presented electronic structure calculations on six lead halide compounds using the modified Becke-Johnson method.  We used the experimental crystal structure of these compounds at room temperature.  We found that by modifying the parameters that characterize the TB-mBJ method, we obtain band gaps that are in excellent agreement with experiment.  Using this new set of parameters, one should be able to predict the electronic structure of phases of these compounds that occur at different temperatures, as well as those of similar compounds obtained by replacing the alkali metal with various organic cations.

\acknowledgments

We gratefully acknowledge support by NSF under grant No. HRD-0932421.


\begin{thebibliography}{10}

\bibitem{ref:Kojima}
Kojima, A.; Teshima, K.; Shirai, Y.; Miyasaka, T. Organometal Halide
  Perovskites as Visible-Light Sensitizers for Photovoltaic Cells.
  \textit{J. Am. Chem. Soc.}  \textbf{2009}, 131, 6050--6051.

\bibitem{ref:Jin}
Jin, H.; Im, J.; Freeman, A.~J. Topological Insulator Phase in Halide
  Perovskite Structures. \textit{Phys. Rev. B}  \textbf{2012}, 86, 121102.

\bibitem{ref:Yang}
Yang, K.; Setyawan, W.; Wang, S.; Nardelli, M.~B.; Curtarolo, S. A Search
  Model for Topological Insulators with High-Throughput Robustness Descriptors.
  \textit{Nature Materials}  \textbf{2012}, 11, 614--619.

\bibitem{ref:Etgar}
Etgar, L.; Gau, P.; Xue, Z.; Peng, Q.; Chandiran, A.~K.; Liu, B.; Nazeeruddin,
  M.~K.; Gr{\"a}tzel, M. Mesoscopic CH$_3$NH$_3$PbI$_3$/TiO$_2$ Heterojunction
  Solar Cells. \textit{J. Am. Chem. Soc}  \textbf{2012}, 134, 17396--17399.

\bibitem{ref:Ball}
Ball, J.~M.; Lee, M.~M.; Hey, A.; Snaith, H.~J. Low-Temperature Processed
  Meso-Superstructured to Thin-Film Perovskite Solar Cells. \textit{Energy Env.
  Sci.}  \textbf{2013}, 6, 1739--1743.

\bibitem{ref:Heo}
Heo, H.~J.; Im, S.~H.; Noh, J.~H.; Mandal, T.~N.; Lim, C.-S.; Chang, J.~A.; Lee, Y.~H.; Kim, H.-j.; Sarkar, A.; Nazeeruddin, Md.~K.; Gr{\"a}tzel, M.; Seok, S.~I. 
  Efficient Inorganic-Organic Hybrid Heterojunction Solar
  Cells Containing Perovskite Compound and Polymeric Hole Conductors.
  \textit{Nature Photonics}  \textbf{2013}, 7, 486--491.

\bibitem{ref:Kim}
Kim, H.-S.; Lee, J.-W.; Yantara, N.; Biox, P.~B.; Kulkarni, S.~A.; Mhaisalker,
  S.; Gr{\"a}tzel, M.; Park, N.-G. High Efficiency Solid-State Sensitized
  Solar Cell-Based on Submicrometer Rutile TiO$_2$ Nanorod and CH$_3$NH$_3$PbI$_3$
  Perovskite Sensitizer. \textit{Nanolett.}  \textbf{2013}, 13, 2412--2417.

\bibitem{ref:Bi}
Bi, D.; Yang, L.; Boschloo, G.; Hagfeldt, A.; Johansson, E. M.~J. Effect of
  Different Hole Transport Materials on Recombination in CH$_3$NH$_3$PbI$_3$
  Perovskite-Sensitized Mesoscopic Solar Cells. \textit{J. Phys. Chem. Lett.}
  \textbf{2013}, 4, 1532--1536.

\bibitem{ref:Cai}
Cai, B.; Xing, Y.; Yang, Z.; Zhang, W.-H.; Qiu, J. High Performance Hybrid
  Solar Cells Sensitized by Organolead Halide Perovskites. \textit{Energy Env.
  Sci.}  \textbf{2013}, 6, 1480--1485.

\bibitem{ref:Eperon}
Eperon, G.~E.; Burlakov, V.~M.; Docampo, P.; Goriely, A.; Snaith, H.~J.
  Morphological Control for High Performance, Solution-Processed Planar
  Heterojunction Perovskite Solar Cells. \textit{Advanced Functional Materials}
   \textbf{2014}, 24, 151--157.

\bibitem{ref:Laban}
Laban, W.~A.; Etgar, L. Depleted Hole Conductor-Free Lead Halide Iodide
  Heterojunction Solar Cells. \textit{Energy Env. Sci.}  \textbf{2014}, 6,
  3249--3253.

\bibitem{ref:Stranks}
Stranks, S.~D.; Eperon, G.~E.; Grancini, G.; Menelaou, C.; Alcocer, M.~J.~P.; Leijtens, T.; Herz, L.~M.; Petrozza, A.; Snaith, H.~J.  
  Electron-Hole Diffusion Lengths Exceeding 1 Micrometer
  in an Organometal Trihalide Perovskite Absorber. \textit{Science}
  \textbf{2013}, 342, 341--344.

\bibitem{ref:Mosconi}
Mosconi, E.; Amat, A.; Nazeeruddin, Md.~K.; Gr{\"a}tzel, M.; De~Angelis, F.
  First-Principles Modeling of Mixed Halide Organometal Perovskites for
  Photovoltaic Applications. \textit{J. Phys. Chem. C}  \textbf{2013}, 117,
  13902--13913.

\bibitem{ref:Wang}
Wang, Y.; Gould, T.; Dobson, J.~F.; Zhang, H.; Yang, H.; Yao, X.; Zhao, H.
  Density Functional Theory Analysis of Structural and Electronic Properties of
  Orthorhombic Perovskite CH$_3$NH$_3$PbI$_3$. \textit{Phys. Chem. Chemical
  Phys.}  \textbf{2014}, 16, 1424--1429.

\bibitem{ref:GW}
Umari, P.; Mosconi, E.; De~Angelis, F. Relativistic GW calculations on
  CH$_3$NH$_3$PbI$_3$ and CH$_3$NH$_3$SnI$_3$ Perovskites for Solar Cell Applications.
  \textit{Scientific Reports}  \textbf{2014}, 4, Article number: 4467.

\bibitem{ref:Lee}
Lee, M.~M.; Teuscher, J.; Miyasaka, T.; Murakami, T.~N.; Snaith, H.~J.
  Efficient Hybrid Solar Cells Based on Meso-Superstructured Organometal Halide
  Perovskites. \textit{Science}  \textbf{2012}, 338, 643--647.

\bibitem{ref:Noh}
Noh, J.~H.; Im, S.~H.; Heo, J.~H.; Mandal, T.~N.; Seok, S.~I. Chemical
  Management for Colorful, Efficient, and Stable Inorganic-Organic Hybrid
  Nanostructured Solar Cells. \textit{Nano Lett.}  \textbf{2013}, 13,
  1764--1769.

\bibitem{ref:Burschka}
Burschka, J.; Pellet, N.; Moon, S.-J.; Humphry-Baker, R.; Gao, P.; Nazeeruddin,
  M.~K.; Gr{\"a}tzel, M. Sequential Deposition as a Route to High-Performance
  Perovskite-Sensitized Solar Cells. \textit{Nature}  \textbf{2013}, 499,
  316--319.

\bibitem{ref:Liu}
Liu, M.; Johnston, M.~B.; Snaith, H.~J. Efficient Planar Heterojunction
  Perovskite Solar Cells by Vapour Deposition. \textit{Nature}  \textbf{2013},
  501, 395--398.

\bibitem{ref:Kohn}
Kohn, W.; Sham, L.~J. Self-Consistent Equations Including Exchange and
  Correlation Effects. \textit{Phys. Rev.}  \textbf{1965}, 140, A1133--A1138.

\bibitem{ref:Perdew}
Perdew, J.~P.; Burke, K.; Ernzerhof, M. Generalized Gradient Approximation
  Made Simple. \textit{Phys. Rev. Lett.}  \textbf{1996}, 77, 3865--3868.

\bibitem{ref:Becke93}
Becke, A.~D. A New Mixing of Hartree-Fock and Local Density-Functional
  Theories. \textit{J. Chem Phys.}  \textbf{1993}, 98, 1372--1377.

\bibitem{ref:Bechstedt}
Bechstedt, F.; Fuchs, F.; Kresse, G. Ab-initio Theory of Semiconductor Band
  Structures: New Developments and Progress. \textit{Phys. Status Solidi B}
  \textbf{2009}, 246, 1877--1892.

\bibitem{ref:Becke06}
Becke, A.~D.; Johnson, E.~R. A Simple Effective Potential for Exchange.
  \textit{J. Chem. Phys.}  \textbf{2006}, 124, 221101.

\bibitem{ref:Tran07}
Tran, F.; Blaha, P.; Schwarz, K. Band Gap Calculations with Becke-Johnson
  Exchange Potential. \textit{J. Phys.: Condens. Matter}  \textbf{2007}, 19,
  196208.

\bibitem{ref:Tran09}
Tran; F.; Blaha, P. Accurate Band Gaps of Semiconductors and Insulators with a
  Semilocal Exchange-Correlation Potential. \textit{Phys. Rev. Lett.}
  \textbf{2009}, 102, 226401.

\bibitem{ref:Becke89}
Becke, A.~D.; Roussel, M.~R. Exchange Holes in Inhomogeneous Systems: A
  Coordinate-Space Model. \textit{Phys. Rev. A}  \textbf{1989}, 39, 3761--3767.

\bibitem{ref:Koller11}
Koller, D.; Tran; F.; Blaha, P. Merits and Limits of the Modified
  Becke-Johnson Exchange Potential. \textit{Phys. Rev. B}  \textbf{2011}, 83,
  195134.

\bibitem{ref:Koller12}
Koller, D.; Tran, F.; Blaha, P. Improving the Modified Becke-Johnson Exchange
  Potential. \textit{Phys. Rev. B}  \textbf{2012}, 85, 155109.

\bibitem{ref:Blaha}
Blaha, P.; Schwarz, K.; Madsen, G. K.~H.; Kvasnicka, D.; Luitz, J.
  WIEN2K: An Augmented Plane Wave and Local Orbitals Program for
  Calculating Crystal Properties, edited by Schwarz, K. Techn. Vienna University of Technology,
   Austria), \textbf{2001}.

\bibitem{ref:Poglitsch}
Poglitsch, A.; Weber, D. Dynamic Disorder in
  Methylammoniumtrihalogenoplumbates (II) Observed by Millimeter-Wave
  Spectroscopy. \textit{J. Chem. Phys}  \textbf{1987}, 87, 6373--6378.

\bibitem{ref:Mashiyama}
Mashiyama, H.; Kurihara, Y.; Azetsu, T. Disordered Cubic Perovskite Structure
  of CH$_3$NH$_3$PbX$_3$ (X=Cl,Br,I). \textit{J. Korean Physical Soc.}
  \textbf{1998}, 32, S156--S158.

\bibitem{ref:Moreira}
Moreira, R.~L.; Dias, A. Comment on ``Prediction of Lattice Constant in Cubic
  Perovskites''. \textit{J. Phys. Chem. Solids}  \textbf{2007}, 68, 1617--1622.

\bibitem{ref:Stoumpos}
Stoumpos, C.~C.; Malliakas, C.~D.; Peters, J.~A.; Liu, Z.; Sebastian, M.; Im,
  J.; Chasapis, T.~C.; Wibowo, A.~C.; Chung, D.~Y.; Freeman, A.~J.; Wessels,
  B.~W.; Kanatzidis, M.~G. Crystal Growth of the Perovskite Semiconductor
  CsPbBr$_3$: A New Material for High-Energy Radiation Detection. \textit{Cryst.
  Growth Des.}  \textbf{2013}, 13, 2722--2727.

\bibitem{ref:Trots}
Trots, D.~M.; Myagkota, S.~V. High-Temperature Structural Evolution of Caesium
  and Rubidium Triiodoplumbates. \textit{J. Phys. Chem. Solids}  \textbf{2008},
  69, 2520--2526.

\bibitem{ref:Baikie}
Baikie, T.; Fang, Y.; Kadro, J.~M.; Schreyer, M.; Wei, F.; Mhaisalkar, S.~G.;
  Graetzel, M.; White, T.~J. Synthesis and Crystal Chemistry of the Hybrid
  Perovskite (CH$_3$NH$_3$)PbI$_3$ for Solid-State Sensitised Solar Cell Applications. \textit{J. Mater.
  Chem. A}  \textbf{2013}, 1, 5628--5641.

\bibitem{ref:Liu13}
Liu, Z.; Peters, J.~A.; Stoumpos, C.~C.; Sebastian, M.; Wessels, B.~W.; Im, J.;
  Freeman, A.~J.; Kanatzidis, M.~G. Heavy Metal Ternary Halides for
  Room-Temperature X-Ray and Gamma-Ray Detection. \textit{Proc. SPIE}
  \textbf{2013}, 8852, 88520A.

\bibitem{ref:Yunakova}
Yunakova, O.~N.; Miloslavskii, V.~K.; Kovalenko, E.~N. Exciton Absorption
  Spectrum of Thin (KI)$_{1-x}$(PbI$_2$)$_x$ films. \textit{Functional
  Materials}  \textbf{2013}, 20, 59--63.

\bibitem{ref:Lambrecht}
Huang, L.~Y.; Lambrecht, W. R.~L. Electronic Band Structure, Phonons, and
  Exciton Binding Energies of Halide Perovskites CsSnCl$_3$, CsSnBr$_3$, and
  CsSnI$_3$. \textit{Phys. Rev. B}  \textbf{2013}, 88, 165203.

\bibitem{ref:gap}
Jiang, H.; G\'{o}mez-Abal, R.~I.; Li, X.-Z.; Meisenbichler, C.; Ambrosch-Draxl, C.;
  Scheffler, M. FHI-gap : A GW Code Based on the All-Electron Augmented Plane
  Wave method. \textit{Computer Phys. Commun.}  \textbf{2013}, 184, 348--366.

\end{thebibliography}
\end{document}